\documentclass[lettersize,journal]{IEEEtran}
\usepackage{amsmath,amsfonts}
\usepackage{algorithmic}
\usepackage{algorithm}
\usepackage{array}
\usepackage{textcomp}
\usepackage{stfloats}
\usepackage{url}
\usepackage{verbatim}
\usepackage{graphicx}
\usepackage{cite}

\usepackage{amsfonts} 
\usepackage{physics} 
\usepackage{algorithmic}  
\usepackage{algorithm} 
\usepackage{graphicx}
\usepackage{color} 
\usepackage{bm}
\usepackage{epstopdf} 

\usepackage{booktabs}  
\usepackage[dvipsnames]{xcolor} 
\usepackage{cite} 
\usepackage{balance}
\usepackage{arydshln} 
\usepackage{mathrsfs} 
\usepackage{cuted} 

\usepackage{multirow} 
\usepackage{makecell} 
\usepackage{float}
\usepackage{subfig}
\usepackage{caption}
\begin{document}
	
	\title{ \LARGE CovNet: Covariance Information-Assisted CSI Feedback for FDD Massive MIMO Systems}
	
	\author{Jialin Zhuang, \textit{Student Member}, \textit{IEEE}, Xuan He, \textit{Graduate Student Member}, \textit{IEEE}, Yafei Wang,  \textit{Graduate Student Member}, \textit{IEEE}, Jiale Liu,  \textit{Student Member}, \textit{IEEE},
		Wenjin Wang, \textit{Member}, \textit{IEEE}
			\thanks{Manuscript received xxx; revised xxx. This work was supported by the National Natural Science Foundation of China under Grant 62341110 and 62371122, the Jiangsu Province Basic Research Project under Grant BK20192002. The associate editor coordinating the review of this article and approving it for publication was xxx. \emph{(Jialin Zhuang and Xuan He contributed equally to this work.)} \emph{(Corresponding author: Wenjin Wang.)}}
		\thanks{Jialin Zhuang and Jiale Liu are with the National Mobile Communications Research Laboratory, Southeast University, Nanjing 210096, China (e-mail: zhuangjl0221@seu.edu.cn; liujiale20021111@seu.edu.cn).}
		\thanks{Xuan He, Yafei Wang, and Wenjin Wang are with the National Mobile Communications Research Laboratory, Southeast University, Nanjing 210096, China, and also with Purple Mountain Laboratories, Nanjing 211100, China (e-mail: hexuan0502@seu.edu.cn; wangyf@seu.edu.cn; wangwj@seu.edu.cn).}
	}


	
	\maketitle
	
	\begin{abstract}
		In this paper, we propose a novel covariance information-assisted channel state information (CSI) feedback scheme for frequency-division duplex (FDD) massive multi-input multi-output (MIMO) systems. Unlike most existing CSI feedback schemes, which rely on instantaneous CSI only, the proposed CovNet leverages CSI covariance information to achieve high-performance CSI reconstruction, primarily consisting of convolutional neural network (CNN) and Transformer architecture. To efficiently utilize covariance information, we propose a covariance information processing procedure and sophisticatedly design the covariance information processing network (CIPN) to further process it. Moreover, the feed-forward network (FFN) in CovNet is designed to jointly leverage the 2D characteristics of the CSI matrix in the angle and delay domains. Simulation results demonstrate that the proposed network effectively leverages covariance information and outperforms the state-of-the-art (SOTA) scheme across the full compression ratio (CR) range.
	\end{abstract}
	

	\bstctlcite{IEEEexample:BSTcontrol}
	\section{Introduction}
	\IEEEPARstart{M}{assive} multiple-input multiple-output (MIMO) is a fundamental technique for the next generation wireless communication system, which has demonstrated significant potential in achieving high spectrum and energy efficiency in wireless communication \cite{RN23}. However, realizing these benefits depends on the accurate acquisition of channel state information (CSI) \cite{RN1}. The channel reciprocity does not hold in frequency-division duplex (FDD) systems, since the uplink and downlink in FDD system are usually separated by more than a coherence frequency bandwidth \cite{RN26}. Therefore, to acquire downlink CSI in FDD system, channel estimation and CSI feedback are required to be performed at the user equipment (UE). The feedback overhead scales linearly with the number of antennas at the base station (BS), which is prohibitive in massive MIMO systems.
	
	The vital need to reduce CSI feedback overhead in FDD massive MIMO systems has motivated various significant researches. In codebook-based CSI feedback methods, the UE feeds back the index of the selected codeword to the BS, and the BS obtains the corresponding codeword based on the received index \cite{RN19}. Nevertheless, the performance of codebook-based CSI feedback accuracy, the overhead of channel codebook, and the complexity are unsatisfactory \cite{RN16}. Compressive sensing (CS) based CSI feedback methods are adopted to reduce the feedback overhead by leveraging the sparsity of CSI in some specific domains \cite{RN2}, such as LASSO \cite{RN20} and block-matching and 3D filtering-approximate message passing (BM3D-AMP) \cite{RN21}. However, CS algorithms typically necessitate multiple iterations for computation and heavily rely on the assumption of channel sparsity, while the CSI matrices may not entirely satisfy this assumption in practical systems. Recently, deep learning (DL) has demonstrated remarkable effectiveness by data-driven patterns in CSI feedback. The first presented neural network, CsiNet \cite{RN3}, exhibits overwhelming superiority over conventional CSI feedback methods, which can be regarded as analogous to an autoencoder. The TransNet proposed in \cite{RN13} further exploits the capabilities of the Transformer architecture \cite{RN12} to achieve state-of-the-art (SOTA) accuracy in CSI recovery.

	Although the aforementioned works produce superior performance under low compression ratio (CR), they show poor performance under high CR as they conduct CSI recovery only based on the instantaneous downlink feedback compressed CSI. Notably, the partial correlation between the uplink and downlink channels still exists in FDD systems. UpAid-FBnet \cite{RN24} and DualNet-ABS \cite{RN25} both utilize uplink information to assist in CSI reconstruction, which leaves out exploring the potential of utilizing statistical information. Additionally, \cite{RN15} reveals that extrapolating downlink covariance information from the estimated uplink covariance information is available. This raises a key question: \emph{How to exploit the downlink statistical information to assist CSI feedback?} In this work, we propose a covariance information-assisted CSI feedback network called CovNet to provide an answer. Specifically, after compressing the CSI at the UE for feedback, the BS exploits a lightweight convolutional neural network (CNN) and Transformer architecture to process the preprocessed covariance information. The processed covariance information vector and the compressed feedback CSI vector are jointly fed into the decoder for the CSI reconstruction. Simulation results show that the proposed CovNet can significantly improve the performance of CSI recovery, especially under high CRs.

	\section{System Model}
	We consider an FDD massive MIMO system operating in orthogonal frequency division multiplexing (OFDM) with $N_{\rm{c}}$ subcarriers, where the UE is equipped with a single antenna, and the BS is equipped with a $N_{\rm{t}}$-antennas uniform linear array (ULA). The downlink received signal at the $i$-th subcarrier is
	\begin{equation}
		\label{deqn_ex1a}
		y_{i}=\tilde{\mathbf{h}}_{i}^{H} \mathbf{s}_{i} x_{i}+z_{i}\text{,}
	\end{equation}
	where $\tilde{\mathbf{h}}_{i}\in \mathbb{C}^{N_{\rm{t}} \times 1}$, $\mathbf{s}_{i}\in \mathbb{C}^{N_{\rm{t}} \times 1}$, $x_{i}$, and $z_{i}$ denotes the channel frequency response vector, precoding vector, transmitted symbol, and additive noise at the $i$-th subcarrier, respectively. The downlink CSI matrix stacked in the spatial-frequency domain is denoted as $\widetilde{\mathbf{H}}=\left[\tilde{\mathbf{h}}_{1},\tilde{\mathbf{h}}_{2}, \cdots, \tilde{\mathbf{h}}_{N_{\rm{c}}}\right]^{H}\in \mathbb{C}^{N_{\rm{c}} \times {N_{\rm{t}}}}$. In such a system, downlink CSI $\widetilde{\mathbf{H}}$ is sent from the UE to the BS through feedback links. Once the BS receives the feedback of $\widetilde{\mathbf{H}}$, it can be utilized to enhance the communication quality, e.g., the design of the precoding vector $\mathbf{s}_{n}$. Due to the extensive deployment of antennas at BS, the total number of parameters to be fed back, i.e., $2N_{\rm{c}}N_{\rm{t}}$, is prohibitive in practical systems.

	In order to reduce feedback overhead in massive MIMO systems, we exploit the sparsity nature of the channel in the angle-delay domain \cite{RN16}. By taking the 2D discrete Fourier transform (DFT), the angle-delay domain channel $\mathbf{H}_{\rm{a}}$ can be expressed as \cite{RN14}
	\begin{equation}
		\label{deqn_ex3a}
		\mathbf{H}_{\rm{a}}=\mathbf{F}_{\rm{d}}\widetilde{\mathbf {H}} \mathbf{F}_{\rm{a}}^{H}\text{,}
	\end{equation}
	where $\mathbf{F}_{\rm{d}} \in \mathbb{C}^{N_{\rm{c}} \times N_{\rm{c}}}$ and $\mathbf{F}_{\rm{a}} \in \mathbb{C}^{N_{\rm{t}} \times N_{\rm{t}}}$ are both DFT matrices. Since the time delay between multiple paths lies within a specifically limited period, most elements in $\mathbf{H}_{\rm{a}}$ are near zero except for the first $N_{\rm{a}}$ rows \cite{RN16}. Hence we denote the truncated matrix as $\mathbf{H} \in \mathbb{C}^{N_{\rm{a}} \times N_{\rm{t}}}$, which consists of the first $N_{\rm{a}}$ rows of $\mathbf{H}_{\rm{a}}$. Then, the total number of feedback parameters can be reduced to $2N_{\rm{a}}N_{\rm{t}}$, which is still not allowed for the limited feedback links.
	
	The CSI feedback problem aims to design compression and recovery approaches for the CSI at the UE and BS to achieve accurate CSI recovery with fewer feedback overhead \cite{RN2,RN3,RN13,RN24,RN25}. The CSI feedback process can be summarized as follows: Once the channel matrix $\mathbf{\widetilde{H}}$ is obtained at the UE, the truncated matrix $\mathbf{H}$ can be acquired by performing 2D DFT in (2). Then, the truncated matrix $\mathbf{H}$ is initially compressed into a low-dimensional signal vector $\mathbf{v}\in\mathbb{C}^{{M} \times{1}}$ at the UE by the compression module, where the number of parameters of $\mathbf{v}$ is significantly smaller than $\mathbf{H}$, i.e., $M\ll2N_{\rm{a}}N_{\rm{t}}$. After that, $\mathbf{v}$ will be transmitted to the BS and used as the input of the recovery module to obtain the reconstructed $\mathbf{H}$. Finally, the channel matrix $\widetilde{\mathbf{H}}$ in the spatial-frequency domain is obtained by operating inverse DFT.

	\section{Covariance Information-Assisted CSI Feedback}
	In this section, we first introduce how to obtain and preprocess covariance information for further processing by covariance information processing network (CIPN). Next, we propose a covariance information-assisted CSI feedback scheme, where the designed CovNet effectively utilizes the covariance information vector provided by the inherent CIPN to assist the decoder for efficient CSI reconstruction.
	\subsection{Covariance Information Preprocessing}
	In the considered system, the BS acquires the uplink CSI covariance information based on the estimated uplink CSI. Specifically, the uplink covariance matrix $\mathbf{C}_{{\rm{ul}},n}\in \mathbb{C}^{N_{\rm{t}} \times {N_{\rm{t}}}}$ with respect to the $n$-th row of the uplink angle-delay domain channel matrix $\mathbf{H}_{\rm{ul}}$ can be calculated by
	\begin{equation}
		\setlength\abovedisplayskip{5pt}
		\mathbf{C}_{{\rm{ul}},n}=\mathbb{E}\left[\mathbf{h} _{n,t}\mathbf{h} _{n,t}^{H}\right]\text{,}\quad n\in\left\{{1,...,N_{\rm{a}}}\right\}\text{,}
		\setlength\belowdisplayskip{5pt}
	\end{equation}
	where $\mathbf{C}_{{\rm{ul}},n} \in \mathbb{C}^{N_{\rm{t}} \times {N_{\rm{t}}}}$ remains constant for relatively long time interval \cite{RN15}, and $\mathbf{h} _{n,t}$ denotes the $n$-th row vector of $\mathbf{H}_{\rm{ul}}$ at the $t$-th period of the sounding reference signal \cite{3gpp.38.211}. Subsequently, downlink CSI covariance information is obtained through extrapolation, based on the assumption that the channel $\textit{Angular Scattering Function}$ is the same for uplink and downlink channel (angular reciprocity) \cite{RN15}, $\mathbf{C}_{{\rm{dl}},n}=f_{\rm{ex}}\left(\mathbf{C}_{{\rm{ul}},n}\right)$, where $f_{\rm{ex}}$ denotes the covariance extrapolation function. Since this work focuses on the utilization of covariance information, we assume perfect uplink channel estimation and covariance extrapolation techniques in the subsequent sections, while the impact of covariance information errors is investigated in Section \uppercase\expandafter{\romannumeral4}. For the simplicity of expression, we refer to the downlink covariance matrix $\mathbf{C}_{{\rm{dl}},n}$ as $\mathbf{C}_{n}$, and define the set of downlink covariance matrices as $\mathcal{D}=\{\mathbf{C}_{n}\}_{n=1}^{N_{\rm{a}}}$.
	
	Furthermore, we consider preprocessing the covariance information. Firstly, the eigenvalue decomposition (EVD) is employed to extract crucial information from high-dimensional covariance matrices. Due to its symmetry, the covariance matrix $\mathbf{C}_{n}$ can be decomposed by EVD as follows 
	\begin{equation}	
		\setlength\abovedisplayskip{5pt}
		\mathbf{C}_{n}=\mathbf{Q} \boldsymbol{\Lambda} \mathbf{Q}^{H}\text{,}
		\setlength\belowdisplayskip{5pt}
	\end{equation}
	where $\mathbf{Q}$ and $\boldsymbol{\Lambda}={\rm{diag}}\left(\lambda_{1,n},...,\lambda_{N_{{\rm{t}}},n}\right)$ are formed by eigenvectors and corresponding eigenvalues of $\mathbf{C}_{n}$, respectively. The first covariance information ${\bar {\bf Q}}\in \mathbb{C}^{N_{\rm{a}} \times N_{\rm{t}}}$ is formed by extracting significant eigenvectors from covariance matrices
	\begin{equation}	
		\setlength\abovedisplayskip{5pt}
		{\bar {\bf Q}}=\left[\mathbf{q}_{1}, \mathbf{q}_{2}, \cdots, \mathbf{q}_{N_{\rm{a}}}\right]^{H}\text{,}
		\setlength\belowdisplayskip{5pt}
	\end{equation}
	where $\mathbf{q}_{n} \in \mathbb{C}^{N_{\rm{t}} \times 1}$ denotes the eigenvector with the largest eigenvalue of the $\mathbf{C}_{n}$. However, ${\bar {\bf Q}}$ only reserves the feature of CSI matrices in eigenspace, which neglects important features in raw space. 
	Accordingly, we retain two matrices of $\mathcal{D}$ for supplying more original information. Specifically, we denote covariance matrices, whose corresponding channel vector has the largest and the second largest power, as $\mathbf{C}_{i_{\rm m1}}$ and $\mathbf{C}_{i_{\rm m2}}$, respectively, i.e., $\Vert \mathbf{h}_{i_{\rm m1}}\Vert_2^{2} \textgreater \Vert \mathbf{h}_{i_{\rm m2}}\Vert_2^{2} \textgreater \Vert \mathbf{h}_{i}\Vert_2^{2}\text{, }\forall i\in\left\{{1,\ldots,N_{\rm{a}}}\right\}\backslash \left\{i_{\rm m1},i_{\rm m2}\right\}$. 
	We concatenate these two matrices along a new dimension, denoted as 
	\begin{equation}	
		\setlength\abovedisplayskip{5pt}
		{\bar {\mathcal{C}}}=\operatorname{concatenate}\left(\mathbf{C}_{i_{\rm m1}}, \mathbf{C}_{i_{\rm m2}}\right)\in \mathbb{C}^{N_{\rm{t}} \times N_{\rm{t}} \times 2}\text{.}
		\setlength\belowdisplayskip{5pt}
	\end{equation}
	${\bar {\mathbf Q}}$ and ${\bar {\mathcal C}}$ are utilized to assist in the subsequent recovery of the compressed CSI.
	
	\subsection{CSI Feedback Scheme}
	Our proposed architecture firstly utilizes the covariance information in the CSI feedback scheme, which is shown in Fig. 1. Concretely,  we design the encoder at the UE to compress the angle-delay domain channel $\mathbf{H}$, i.e.,
	\begin{equation}	
		\label{deqn_ex5a}
		\mathbf{v}= f_{\rm{e}}\left(\mathbf{H} ; \Theta_{\rm{e}}\right)\text{,}
	\end{equation}
	where $\mathbf{v}$ is the compressed $M$-dimensional information vector, $\Theta_{\rm{e}}$ denotes the parameters of the designed encoder, and the CR can be calculated by $\rm{CR}=2 N_{\rm{a}} N_{\rm{t}}/M$.
	\begin{figure}[H]
		\captionsetup{justification=raggedright,singlelinecheck=false}
		\centering
		\includegraphics[width=3in]{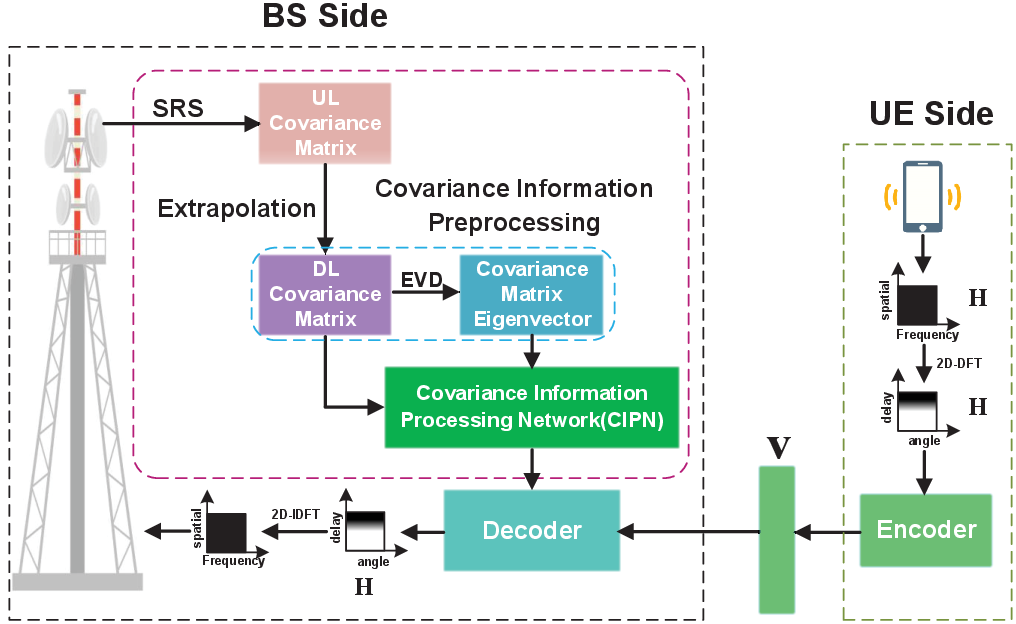}
		\caption{The architecture of covariance information-assisted CSI feedback scheme.}
		\label{fig_2}
	\end{figure} 
	\begin{figure*}[b]
		\captionsetup{justification=raggedright,singlelinecheck=false}
		\centering
		\includegraphics[width=5.72in]{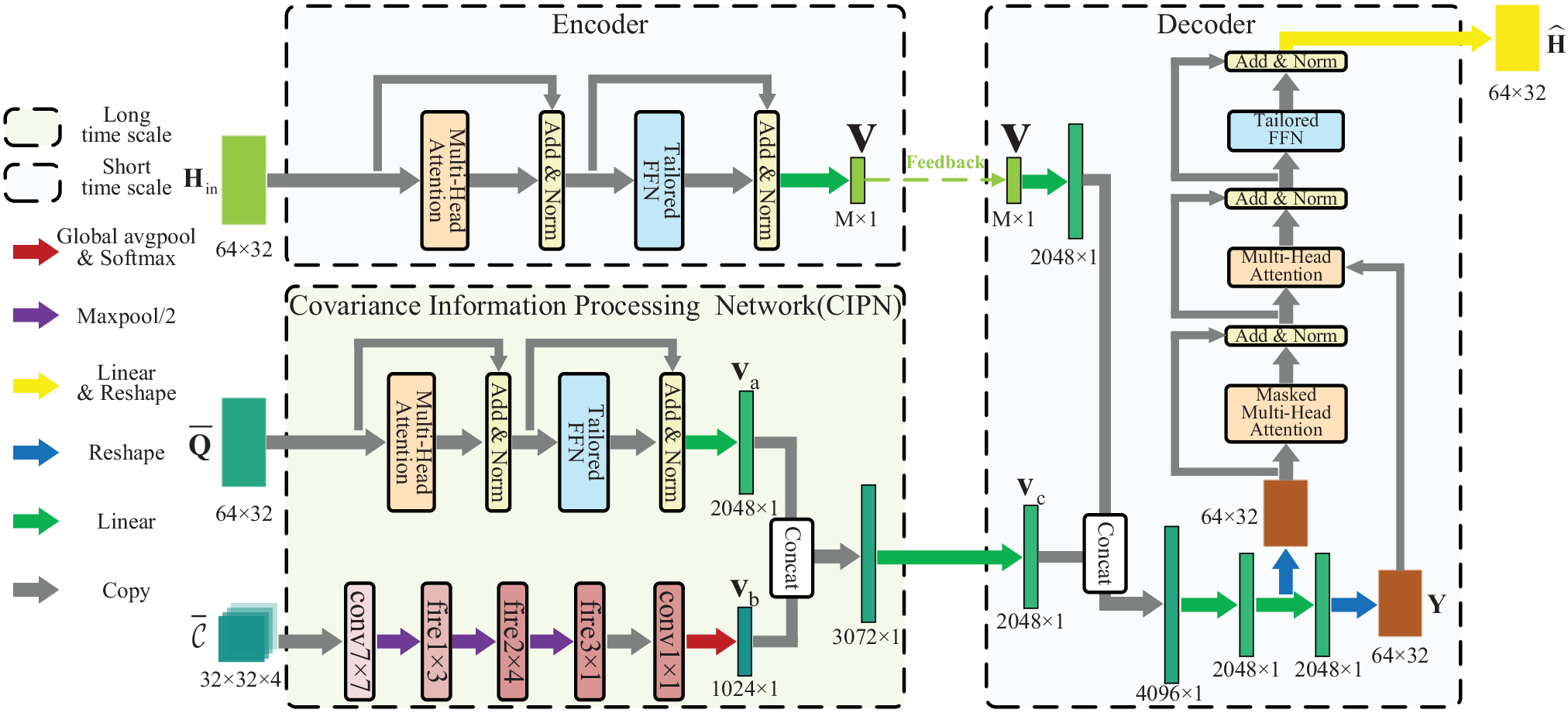}
		\caption{The architecture of CovNet. We assume $N_{\rm{a}}=N_{\rm{c}}=32$.}
		\label{fig_1}
	\end{figure*} 
	Assuming the channel covariance matrix is perfectly obtained at the BS, then the covariance information vector applicable to assist CSI feedback can acquired through elaborate transformation on the  preprocessed covariance information, which is designed as
	\begin{equation}	
		\label{deqn_ex5a}
		\begin{aligned}
			\mathbf{v}_{\rm{c}}= f_{\rm{CIPN}}\left({\bar {\bf Q}}\text{, } {\bar {\mathcal{C}}}; \Theta_{\rm{CIPN}}\right)\text{,}
		\end{aligned}
	\end{equation}
	where $\mathbf{v}_{\rm{c}}$ is the covariance information vector extracted through the CIPN, and $\Theta_{\rm{CIPN}}$ denotes the parameters of the designed CIPN.
	Additionally, with the assistance of covariance information, we design the decoder from the codeword to the CSI matrix $\mathbf{H}$, i.e.,
	\begin{equation}	
		\label{deqn_ex5a}
		{\hat{\mathbf{H}}}= f_{\rm{d}}\left(\mathbf{v} \text{, }\mathbf{v}_{\rm{c}}
		; \Theta_{\rm{d}}\right)\text{,}
	\end{equation}
	where $\Theta_{\rm{d}}$ denotes the parameters of the designed decoder. The process of covariance information-assisted CSI feedback architecture can be formulated as (4)-(9) sequentially.

	\subsection{CovNet for CSI Feedback}
	Considering that $\mathbf{H}$ comprises delay domain vectors, the encoder is tailored based on Transformer architecture \cite{RN12}, which can achieve outstanding performance across various tasks, especially when processing multiple vectorized data streams. In the first part of CIPN, we also employ a Transformer-based network, which is the same as the encoder, due to the structural similarities between ${\bar {\bf Q}}$ and $\mathbf{H}$. In addition, considering the spatial structure of ${\bar {\mathcal{C}}}$ is similar to 2D images, we integrate CNN in the second part, owing to its excellent image data processing capability \cite{RN16}. We exploit the CNN and Transformer architecture for the encoder, CIPN, and decoder in CovNet, which can exploit the additional covariance information and learn features in the angle-delay domain. As shown in Fig. 2, CovNet is primarily composed of the encoder at the UE and CIPN and decoder at the BS, with the real and imaginary parts of ${\bar {\bf Q}}$, ${\bar {\mathcal{C}}}$, and $\mathbf{H}$ being its input.
	
	Specifically, the input of the encoder $\mathbf{H}_{\rm{in}}\in \mathbb{R}^{2 N_{\rm{a}} \times N_{\rm{t}}}$ is comprised of the real and imaginary parts of $\mathbf{H}$, which is projected  to  different representation subspaces by $ N_{\rm{h}} $ sets of matrices. The Transformer architecture calculates the intermediate variable $\mathbf{A}_{n}$ using the following formula:

	\begin{equation}	
		\setlength\abovedisplayskip{5pt}
		\begin{aligned}
			&\quad{\mathbf{A}}_{n}={\rm{Attention}}\left(\mathbf{H}_{\rm{in}} \mathbf{W}_{n}^{{Q}}, \mathbf{H}_{\rm{in}} \mathbf{W}_{n}^{{K}}, \mathbf{H}_{\rm{in}} \mathbf{W}_{n}^{{V}}\right)\text{,}\\
			&{\rm{Attention}}(\mathbf{Q}_{n}, \mathbf{K}_{n},\mathbf {V}_{n})={\rm{softmax}}\left(\frac{\mathbf{Q}_{n} \mathbf{K}_{n}^{T}}{\sqrt{{d_{k}}}}\right) \mathbf{V}_{n}\text{,}
		\end{aligned}
		\setlength\belowdisplayskip{5pt}
	\end{equation}
	where $\mathbf{W}_{n}^{{Q}}\text{ and }\mathbf{W}_{n}^{{K}}\in \mathbb{R}^{d \times d_{k}}\text{, }\mathbf{W}_{n}^{{V}}\in \mathbb{R}^{d \times d_{v}}$ are learnable parameter matrices, $\mathbf{Q}_{n}=\mathbf{H}_{\rm{in}} \mathbf{W}_{n}^{{Q}}$, $\mathbf{K}_{n}=\mathbf{H}_{\rm{in}} \mathbf{W}_{n}^{{K}}\in \mathbb{R}^{2N_{\rm{a}} \times d_{k}}$, and $\mathbf{V}_{n}=\mathbf{H}_{\rm{in}} \mathbf{W}_{n}^{{V}}\in \mathbb{R}^{2N_{\rm{a}} \times d_{v}}$ represent query matrix, key matrix, and value matrix, respectively, $n=\left\{1,\dots,N_{\rm{h}}\right\}$, $d_{k}=d_{v}=d/N_{\rm{h}}$, and $d=N_{\rm{t}}$ represents the dimension of the input delay domain vectors. After that, the output of the multihead attention layer can be obtained by multiplying the concatenation of self-attention heads, that is
	\begin{equation}	
		\setlength\abovedisplayskip{5pt}
		\mathbf{A}=\left[{\mathbf{A}}_{\rm{1}},\dots,{\mathbf{A}}_{N_{\rm{h}}}\right]\mathbf{W}^{\rm{O}}\in \mathbb{R}^{2N_{\rm{a}} \times d}\text{, }
		\setlength\belowdisplayskip{5pt}
	\end{equation}
	where $\mathbf{W}^{\rm{O}} \in  \mathbb{R}^{d \times d}$ is the learnable parameter matrix. The features of downlink CSI are exploited by the multihead attention mechanism, which are mapped to the codeword $\mathbf{v}$ of length $M$ by the final linear layer in the CovNet encoder.
	
	As shown in Fig. 2, the CIPN consists of two parts to process ${\bar {\bf Q}}$ and ${\bar {\mathcal{C}}}$ separately. Specifically, ${\bar {\mathcal{C}}}$ is manipulated by a lightweight CNN called SqueezeNet \cite{RN17}, which comprises convolutional layers, pooling layers, and Fire modules. The covariance feature vectors, i.e., $\mathbf{v}_{\rm{a}}$ and $\mathbf{v}_{\rm{b}}$, are obtained by Transformer-based network and SqueezeNet in CIPN, respectively. Subsequently, these vectors are concatenated to generate the final covariance information vector $\mathbf{v}_{\rm{c}}$ through the last linear layer in the CIPN, which has the same length as the downlink CSI $\mathbf{H}$. Notably, since $\mathbf{C}_{n}$ is long-term invariant, the complexity of the preprocessing procedure and the CIPN can be considered negligible. This offers significant benefits for the practical implementation of CovNet, as the encoder and decoder primarily determine the overall complexity.
	
	Similar to the encoder structure, the masked multihead attention layer is additionally incorporated in the decoder. Unlike existing works, with the assistance of covariance information vector $\mathbf{v}_{\rm{c}}$, the input of the attention module in the CovNet decoder, denoted as $\mathbf{Y}$, can be obtained by concatenating $\mathbf{v}_{\rm{c}}$ and $\mathbf{v}$ with the linear transformation. Consequently, $\mathbf{Y}$ and the masked multihead attention layer outputs are simultaneously sent to the multihead attention layers. The latter is used to construct the query matrix, while the former is used to construct the key and value matrices. Finally, the output of the multihead attention module is fed into a linear layer with a reshape operation to obtain the estimated ${\hat{\mathbf{H}}}$.
	
	\begin{figure}[H]
		\captionsetup{justification=raggedright,singlelinecheck=false}
		\centering
		\includegraphics[width=3.5in]{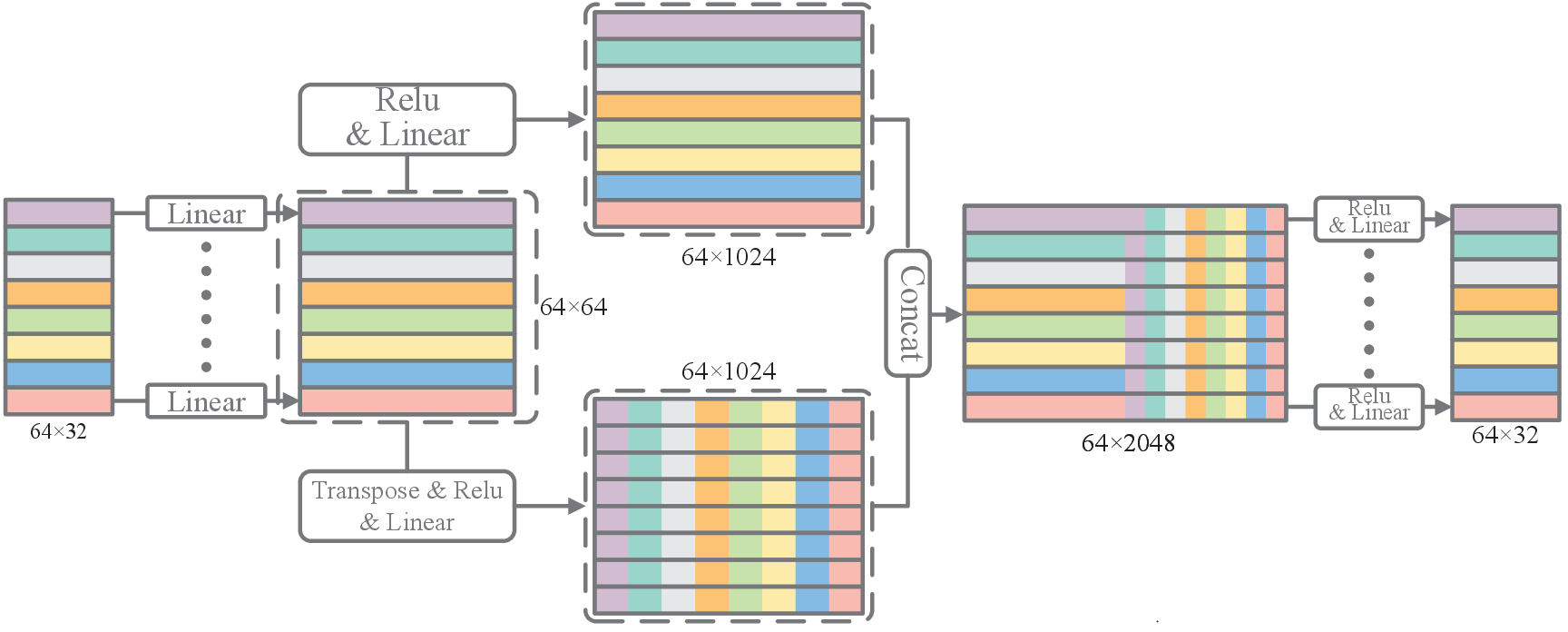}
		\caption{The tailored FFN in CovNet. We assume $N_{\rm{a}}=N_{\rm{c}}=32$.}
	\end{figure} 
	Furthermore, the feed-forward network (FFN) in CovNet is sophisticatedly designed and distinct from the conventional FFN in Transformer architecture. Fig. 3 illustrates the tailored FFN in CovNet, which performs feedforward processing independently in both the angle domain and delay domain to achieve the interaction of two-dimensional features.
	
	\section{Simulation Results}
	We generate two types of channel matrices by COST2100 channel model \cite{RN18} for training and testing: (a) indoor channel with 5.1 GHz uplink and 5.3 GHz downlink bands; (b) semi-urban outdoor channel with 260 MHz uplink and 300 MHz downlink bands. The ULA and OFDM operating in the FDD mode are employed at the BS with $N_{\rm{t}}=32$ antennas and $N_{\rm{c}}=1024$ subcarriers, respectively. When acquiring the angle-delay domain channel matrix $\mathbf{H}$, we only retain the first $N_{\rm{a}}=32$ rows of the channel matrix. Notably, the proposed CovNet can be flexibly extended to larger systems with more subcarriers or antennas. The number of attention heads is set as $N_{\rm{h}}=2$. The datasets are randomly divided into training and testing sets, with 100,000 and 20,000 samples, respectively. The CovNet is trained by an end-to-end strategy on a single NVIDIA 3050 GPU, where the ADAM optimizer is applied with a fixed learning rate $1\times10^{-4}$ to minimize the mean square error (MSE) loss function. The batch size and epoch are set to 200 and 400. The presented CovNet is compared with CsiNet \cite{RN3},  UpAid-FBnet \cite{RN24}, DualNet-ABS \cite{RN25}, and the SOTA TransNet \cite{RN13}, the metric to evaluate the models is chosen as normalized mean-squared error (NMSE) in the angle-delay domain, i.e., $\mathrm{NMSE}=\mathrm{E}\left\{\|\mathbf{H}-\hat{\mathbf{H}}\|_{2}^{2}/{\|\mathbf{H}\|_{2}^{2}}\right\}$, where $\mathbf{H}$ and $\hat{\mathbf{H}}$ are true and reconstruction downlink CSI in the angle-delay domain, respectively.

	We comprehensively analyze the downlink CSI feedback accuracy achieved by CsiNet, TransNet, DualNet-ABS, Upaid-FBnet, and CovNet under four CRs: 32, 64, 128, and 256. To evaluate the performance of networks, particularly under high CRs, we choose 256 as the maximum CR instead of 64. Table I provides the NMSE results for indoor and outdoor environments and computation complexity measured by floating point operations (FLOPs). CovNet and TransNet utilize attention mechanisms, resulting in higher FLOPs but significant performance gains. Leveraging covariance information in addition to compressed CSI vectors, CovNet outperforms CsiNet and TransNet across a wide CR range, which implies that the designed CIPN effectively processes the covariance information. In indoor environments, experiments shown in Table I reveal that our proposed scheme outperforms TransNet by significant margins: 3.91 dB, 3.81 dB, 3.56 dB, and 3.31 dB when CR$=$32, 64, 128, and 256, respectively. As shown in Table I, this performance benefit extends to the more complicated outdoor scenarios, implying its notable robustness and generalization capability across diverse environments. Significantly, the performance of CovNet is slightly impacted by increasing CRs, as the covariance information obtained at BS does not require compression. In contrast, both TransNet and CsiNet struggle to maintain performance under high CRs due to the limitation of compressed codeword length. Moreover, both DualNet-ABS and Upaid-FBnet utilize instantaneous uplink information, which result in better performance under high CRs compared to TransNet.
	
	However, both DualNet-ABS and UpAid-FBnet rely on instantaneous channel information, where the significant frequency band differences between uplink and downlink limit the effectiveness of uplink assistance. In contrast, CovNet utilizes statistical information, which exhibits smaller discrepancies between the uplink and downlink bands \cite{RN15}. CovNet also enhances robustness by preprocessing covariance information through feature compression and denoising. Additionally, by integrating CNN and Transformer architectures, CovNet more accurately extracts substantial features inherent to channels, which can also be demonstrated by the superior performance of TransNet compared to these uplink-assisted networks under low CR.
	
	Furthermore, we modify the proposed CovNet: The tailored FFN and CIPN in CovNet are replaced with conventional FFN and CNN. It can be seen that CovNet outperforms the modified CovNet, demonstrating the effectiveness of the tailored FFN and proposed CIPN in improving CSI recovery accuracy. Overall, the simulations show the superiority of CovNet in achieving high accuracy especially under high CRs.
	
	\begin{table*}[htbp]
		\centering
		\caption{NMSE (dB) and FLOPs Comparison Between CovNet and Other Methods}
		\resizebox{\textwidth}{!}{
			\begin{tabular}{cccccccccccccc}
				\hline
				CR &  \multicolumn{3}{c}{32} &  \multicolumn{3}{c}{64} &  \multicolumn{3}{c}{128} &  \multicolumn{3}{c}{256} & \\
				\hline
				\multirow{2}{*}{Methods} & \multirow{2}{*}{FLOPs} & \multicolumn{1}{c}{NMSE} & \multicolumn{1}{c}{NMSE} & \multirow{2}{*}{FLOPs} & \multicolumn{1}{c}{NMSE} & \multicolumn{1}{c}{NMSE} & \multirow{2}{*}{FLOPs} & \multicolumn{1}{c}{NMSE} & \multicolumn{1}{c}{NMSE} & \multirow{2}{*}{FLOPs} & \multicolumn{1}{c}{NMSE} & \multicolumn{1}{c}{NMSE} &  \\
				&  & \multicolumn{1}{c}{indoor} & \multicolumn{1}{c}{outdoor} &  & \multicolumn{1}{c}{indoor} & \multicolumn{1}{c}{outdoor} &  & \multicolumn{1}{c}{indoor} & \multicolumn{1}{c}{outdoor} &  & \multicolumn{1}{c}{indoor} & \multicolumn{1}{c}{outdoor}  \\
				\hline
				CovNet & \textcolor{red}{35.15M} & \textbf{-5.68} & \textbf{-4.93} &\textcolor{red}{35.02M} & \textbf{-4.66} & \textbf{-4.18} & \textcolor{red}{34.96M} & \textbf{-4.00} & \textbf{-3.73} & \textcolor{red}{34.92M} & \textbf{-3.49} & \textbf{-3.36}  \\
				Modified CovNet & 26.24M & -4.74 & -4.39 & 26.11M & -2.89 & -4.03 & 26.04M & -1.86& -2.84 & 26.01M & -1.25 & -1.72  \\
				Upaid-FBnet & 16.97M & -2.17 & -1.26 & 16.88M & -1.49 & -0.93 & 16.83M & -1.12 & -0.75 & 16.80M & -0.84 & -0.66  \\
				DualNet-ABS & 4.63M & -1.42 & -1.12 & 4.50M & -0.93 & -0.23 & 4.43M & -0.65 & -0.12 & 4.40M & -0.54 & -0.10  \\
				TransNet & \textcolor{red}{35.50M} & -1.77 & -2.40& \textcolor{red}{35.37M} & -0.85 & -1.20 & \textcolor{red}{35.31M} & -0.44 & -0.42 & \textcolor{red}{35.27M} & -0.28 & -0.22  \\
				CsiNet & 3.69M & -0.81 & -0.15 & 3.56M & -0.31 & -0.08 & 3.49M & -0.23 & -0.03 & 3.46M & -0.04 & -0.01 \\

				\hline
		\end{tabular}}
	\end{table*}
	\begin{figure}[h]
		\captionsetup{justification=raggedright,singlelinecheck=false}
		\centering
		\includegraphics[width=3in]{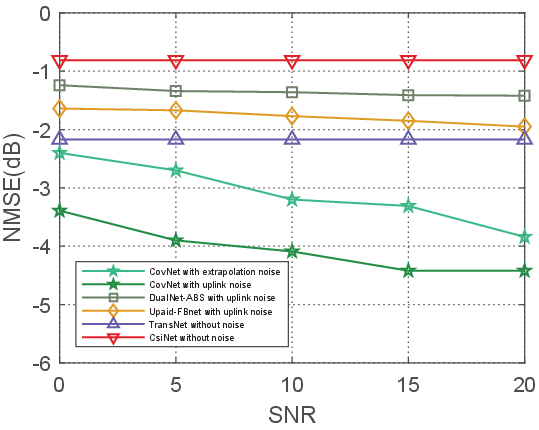}
		\caption{NMSE comparison under different noise powers.}
	\end{figure} 
	While the aforementioned simulations assume the ideal acquisition of covariance information, practical scenarios often involve inaccuracies in estimation. To this end, White Gaussian Noise is added to the covariance matrices and uplink channel matrix obtained at the BS, i.e., ${\tilde {\bf C}_{n}}={\bf C}_{n}+{\bf E}\text{, }{\tilde {\bf H}_{\rm ul}}={\bf H}_{\rm ul}+{\bf E}$. As depicted in Fig. 4, we test different levels of errors in terms of signal-to-noise ratio (SNR). Notably, only the three networks processing uplink information are affected, as the errors considered are unrelated to downlink acquisition and feedback. Notably, with the impact of imperfect covariance information estimation, our network outperforms both DualNet-ABS and Upaid-FBnet with the same SNR conditions. Remarkably, with uplink  $\rm{SNR}=0 \text{ dB}$, the proposed CovNet still outperforms TransNet and CsiNet by 1.62 dB and 2.58 dB, respectively, which infers its robustness in confronting imperfect covariance information acquisition. The robustness of CovNet stems from its exceptional ability to learn in high-dimensional feature spaces and effectively fit complex, noisy data due to its large number of parameters.
	\section{Conclusion}
	In this letter, we proposed a DL-based CSI feedback scheme called CovNet, which leveraged covariance information for CSI recovery. In addition, we first put forward the covariance information preprocessing that facilitated subsequent CIPN design. Based on the latter, CovNet effectively utilized covariance information. The simulation results showed that CovNet outperformed the SOTA CSI feedback scheme in the full CR range with and without noise, particularly showcasing its superiority at high CRs.

	\bibliographystyle{IEEEtran}
	\bibliography{IEEEabrv,IEEEfull}
\end{document}